\documentclass[aps,pre,twocolumn,groupedaddress]{revtex4}

\usepackage{graphicx}

\begin{document}


\title{Random Close Packing of Disks and Spheres in Confined Geometries}


\author{Kenneth W. Desmond and Eric R. Weeks}
\affiliation{Department of Physics, Emory University, Atlanta GA
30322}


\date{\today}

\begin{abstract}
Studies of random close packing of spheres have advanced our knowledge
about the structure of systems such as liquids, glasses,
emulsions, granular media, and amorphous solids. When 
these systems are confined their structural properties
change. To understand these changes we study random close
packing in finite-sized confined systems, in both two and
three dimensions.  Each packing consists of a 50-50 binary
mixture with particle size ratio 1.4.  The presence of confining walls
significantly lowers the overall maximum area fraction (or volume
fraction in three dimensions).  A simple model is presented which
quantifies the reduction in packing due to wall-induced
structure.  This wall-induced structure decays rapidly away
from the wall, with characteristic length scales comparable
to the small particle diameter.
\end{abstract}


\maketitle


\section{Introduction \label{section:Introduction}}

In 1611 Kepler conjectured that the most efficient packing of
spheres is the face center cubic packing ($fcc$), with packing
fraction $\phi_{fcc} = \pi/\sqrt{18}$~\cite{ZamponiNature2008}.
In 1831, Gauss provided a partial proof to this
conjecture, and more recently Hales presented a more
complete proof~\cite{Hales2005} that is still being
validated~\cite{ZamponiNature2008}.  However, the $fcc$
packing is highly ordered, and in many cases random packings
are of interest, as they are easy to create.  For example, a
large range of packing fractions have been found for granular
particles with a minimum mechanically stable volume fraction
$\phi_{rlp} \approx 0.55$~\cite{Scott1969, BerrymanPRA1983,
OnodaPRL1990}, termed random loose packing.  The maximum volume
fraction for a randomly packed 3D system is $\phi_{rcp} \approx
0.64$, termed random close packing ($rcp$)
\cite{BernelNature1960,
Scott1969, BerrymanPRA1983, OHernPRL2002, XuPRE2005}.

The earliest known experimental study on the density of $rcp$
was performed by Bernal and Mason~\cite{BernelNature1960}. In
their experiment they repeatedly shook and compressed
a rubber balloon full of spheres for sufficiently long
enough time to reach a very dense state at $\phi_{rcp} =
0.637$.  This result is sensitive to the preparation method,
although the final volume fractions found are always close
to 0.64 \cite{TorquatoPRL2000}.  Computational studies
are especially sensitive to protocol, with $\phi_{rcp}$
ranging between 0.64 and 0.68~\cite{JodreyPRA1985,
TobochnikChemPhys1988}. A broader range of mechanically
stable packing fractions can be obtained by consider packings
consisting of a polydisperse mixture of spheres or packings with
non-spherical particles~\cite{KansalChemPhys2002, Alraoush2007,
Lochmann2006, Richard2001, Brouwers2006, Donev2004, Okubo2004,
DesmondPRE2006, JiaoPRL2008}.  Some theoretical progress
explaining random close packing has occurred \cite{Edwards1994,
Radin2008, JalaliChemPhys2004, SongNature2008}, although this
is far from complete \cite{ZamponiNature2008}.  All of these
studies are relevant to a wide range of problems including
the structure of living cells~\cite{TorquatoPRL2000},
liquids~\cite{BernelNature1960}, granular
media~\cite{Edwards1994, Radin2008}, emulsions~\cite{Pal2008},
glasses~\cite{Lois2008}, and amorphous solids~\cite{Zallen1983}.

While all of the above studies focus on infinite systems,
real systems have boundaries and often these boundaries
are important.  Furthermore, in many cases samples are
confined between closely spaced boundaries, with the confining
size being only a few characteristic particle sizes across
\cite{Alcoutlabi2005}.  For example, when a liquid is confined
its structure is dramatically changed with particles forming
layers near the wall, which ultimately affects the properties
of the liquid~\cite{GalloChemPhys2000, GranickScience1991,
Henderson2007, MittalChemPhys2007, MittalChemPhys2007_2,
MittalPRL2008}. The shearing of confined dense colloidal
suspensions shows the emergence of new structures not seen
before \cite{CohenPRL2004}. The flow of granular media through
hoppers~\cite{To2002, Zuriguel2003} or suspensions through
constricted micro- and nanofluidic devices~\cite{Redner2000,
Fuller2002, Sharp2005, Whitesides2006} can jam and clog,
costing time and money.  Some studies examined the packing
of granular particles in narrow silos, focusing on stresses
between particles and the walls \cite{Claudin2000, Marconi2000,
Claudin1997, Landry2004}.  Other studies noted that the particle
packing in silos is layered near the walls \cite{LaundryPRE2003,
SeidlerPRE2000}.  However, the influence of boundaries on
$\phi_{rcp}$ has not been studied previously.

We present computer simulated $rcp$ packings in confined
geometries.  In particular, we study binary mixtures to
prevent wall-induced crystallization \cite{Teng2003, Murray1998, Nemeth1999}.  We create two-dimensional (2D) and
three-dimensional (3D) packings between two parallel walls,
with periodic boundary conditions in the other directions.
Confinement significantly modifies the $rcp$ states, with lowered
values for $\phi_{rcp}$ reflecting an inefficient packing near
the walls.  This inefficient packing persists
several particle diameters away from the wall, although its
dominant effects are only within 1-2 diameters.

These results will be useful for understanding many other
confined systems.  For example, many experiments study how
confinement modifies the glass transition; samples which
have a well-characterized glass transition in large samples
show markedly different properties when confined to small
samples \cite{NugentPRL2007, Morineau2002, Schuller1994,
Jackson1991, Kim2000, Thompson1992, Nemeth1999, Scheidler2002,
Alcoutlabi2005}. However, it's not clear if these changes are due
to boundaries or finite size effects \cite{He2007}.  Our results show that
boundaries significantly modify the packing, which may in turn
modify behavior of these confined molecular systems
\cite{Alcoutlabi2005}.

The manuscript is organized as follows. Section
\ref{section:Method} outlines the algorithm we use to
generate confined $rcp$ states. Section~\ref{section:Results}
shows how the total packing fraction, particle number
density, and local order of confined $rcp$ states change
with confining thickness and distance from the confining
boundary. Finally, Sec.~\ref{section:Discussion} 
provides a simple model that predicts the
packing fraction dependence with confinement.

\section{Method \label{section:Method}}

Our aim is to quantify how a confining boundary alters the
structure of randomly closed packed ($rcp$) disks in 2D and
spheres in 3D, and in particular to study how this depends on 
the narrowest dimension.  This section presents our algorithm for
2D packings first, and then briefly discusses differences for the
3D algorithm.

In 2D, our system consists of a binary mixture of disks
containing an equal number $N/2$ large disks of diameter $d_l$
and small disks of diameter $d_s$ with size ratio $\sigma =
d_l/d_s = 1.4$. For each configuration, disks are packed into a
box of dimensions $L_x$ by $L_y$, with a periodic boundary along
the $x$-direction and a fixed hard boundary (a wall) along the
$y$-direction.

\begin{figure}
\includegraphics[width=3.4in]{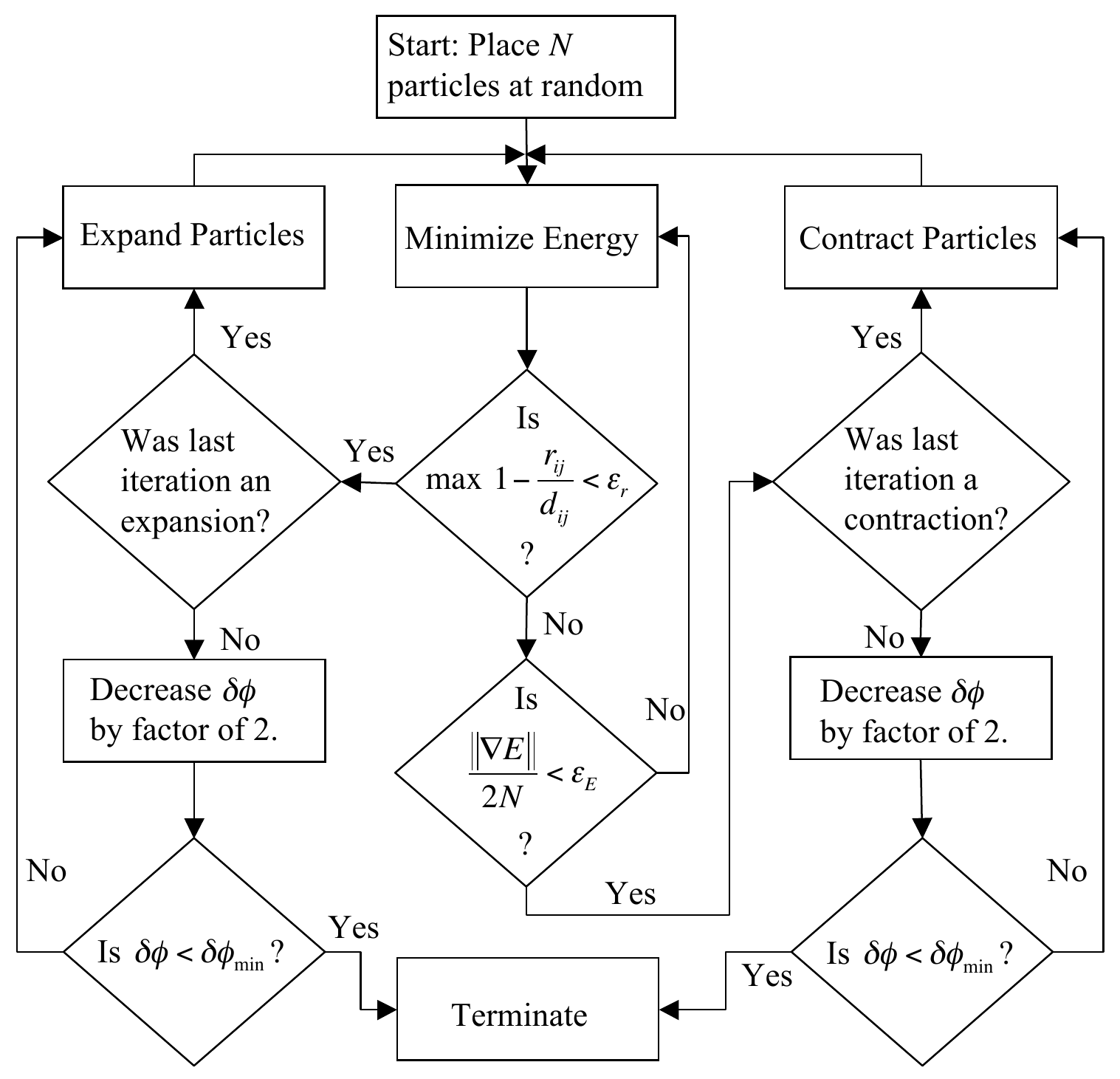}
\caption{A flow chart outlining our algorithm for computing $rcp$ configurations.}
\label{fig:FlowChart}
\end{figure}

Each configuration is generated using a method
adapted from Xu et al.~\cite{XuPRE2005} which is an extension
of a method proposed by Clarke and Wiley~\cite{Clarke1987}.  This
method is briefly summarized in Fig.~\ref{fig:FlowChart}.
Infinitesimal particles are placed in the system, gradually
expanded and moved at each step to prevent particles from
overlapping.  When a final state is found such that particles can
no longer be expanded without necessitating overlap, the
algorithm terminates.  Near the conclusion of the algorithm, we
alternate between expansion and contraction steps to accurately
determine the $rcp$ state.

In particular, while the final state found is consistent with
hard particles (no overlaps allowed), the algorithm uses a soft
potential at intermediate steps \cite{XuPRE2005}, given by
\begin{equation}
V(r_{ij}) = \frac{\epsilon}{2}\left(1- r_{ij}/d_{ij}\right)^2\Theta{}\left(1- r_{ij}/d_{ij} \right),
\label{potential}
\end{equation}
where $r_{ij}$ is the center to center distance between two
disk $i$ and $j$, $\epsilon$ is a characteristic energy scale
($\epsilon = 1$ for our simulations), $d_{ij} = (d_i +d_j)/2$,
and $\Theta{}\left(1 - r_{ij}/d_{ij} \right)$ is the Heaviside
function making $V$ nonzero for $r_{ij} < d_{ij}$. Simulations
begin by randomly placing disks within a box of desired
dimensions and boundary conditions with the initial diameters
chosen such that $\phi_{initial} \ll \phi_{rcp}$.   In the
initial state particles do not overlap and the total energy
$E = 0$.

Next all disk diameters are slowly expanded subject to the fixed
size ratio $\sigma = 1.4$ and $\phi$ changing by $\delta\phi$
per iteration; we start with $\delta\phi=10^{-3}$.  After each
expansion step, we check if any disks overlap, by checking
the condition $1 - r_{ij}/d_{ij} > \epsilon_r = 10^{-5}$ for
each particle pair.  Below this limit, we assume the overlap
is negligible.  If any particles do overlap ($E>0$), we use
the non-linear conjugate gradient method~\cite{Nocedal1999}
to decrease the total energy by adjusting the position of
disks so they no longer overlap ($E = 0$).  In practice, one
energy minimization step does not guarantee we have reached
a minimum within the desired numerical precision.  Thus this
step can be repeated to further reduce the energy if $E > 0$.
We judge that we have reached a nonzero local minimum if the
condition $||\nabla E||/(2N) < \epsilon_E=10^{-7}$ is found,
where $||\nabla E||$ is the magnitude of the gradient of $E$.
Physically speaking this is the average force per particle,
and the threshold value ($10^{-7}$) leads to consistent results.

If we have such a state with $E>0$, this is not an $rcp$ state as
particles overlap.  Thus we switch and now slowly contract the
particles until we find a state where particles again no longer
overlap (within the allowed tolerance).  At that point, we once
again begin expansion.  Each time we switch between expansion and
contraction, we decrease $\delta\phi$ by a factor of 2.  Thus,
these alternating cycles allow us to find an $rcp$ state of
non-overlapping particles (within the specified tolerance) and
determine $\phi_{rcp}$ to high accuracy.  We terminate our
algorithm when $\delta\phi < \delta\phi_{\rm min} = 10^{-6}$.
In practice, we have tested a variety of values for the
thresholds $\epsilon_r, \epsilon_E,$ and $\delta\phi_{\rm min}$
and find that our values guarantee reproducible results as well
as reasonably fast computations.
Our algorithm gave an average packing fraction
of  $\phi_{rcp} = 0.8420 \pm 0.0005$ for 40 simulated $rcp$
states containing 10,000 particles with periodic boundary
conditions along both directions. Our value of  $\phi_{rcp}$
is in agreement with that found by Xu et al.~\cite{XuPRE2005}.

\begin{figure}
\includegraphics[width=3.4in]{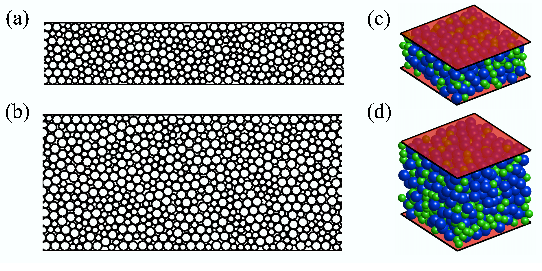}
\caption{(Color online). Illustrations of 2D and 3D configurations generated using the algorithm described in Sec.~\ref{section:Method}. (a) 2D configuration for $h = 10$. (b) 2D configuration with $h = 20$. (c) 3D configuration with $h = 5$ where blue (dark gray) represents big particles and green (light gray) represents small particles. (d) 3D configuration with $h = 10$.}
\label{fig:Configs}
\end{figure}

The above procedure is essentially the same as
Ref.~\cite{XuPRE2005}; we modify this to include the influence of
the boundaries.  To add in the wall, we create image particles
reflected about the position of the wall; thus particles interact
with the wall using the same potential, Eqn.~\ref{potential}.

Additionally, we wish to generate packings with pre-specified
values for the final confining height $h = L_y/d_s$.
(This allows us to create multiple $rcp$ configurations with
the same $h$.)  We impose $h$ by affinely scaling the system
after each step, so that the upper boundary is adjusted by $L_y
= hd_s$ and each disk's $y$-coordinate is multiplied by the
ratio $L_{y, i+1}/L_{y, i}$, where $L_{y, i}$ and $L_{y, i+1}$
are the confining widths between two consecutive iterations.
Thus while $d_s$ gradually increases over the course of
the simulation, $L_y$ increases proportionally so that the
nondimensional ratio $h$ is specified and constant.  Some
examples of our final $rcp$ states are shown in
Fig.~\ref{fig:Configs}.

To ensure we will have no finite size effects in the periodic
direction, we examined $\phi_{rcp}$ for different $h$ and
$L_x$, and found $\phi_{rcp}(h)$ to be independent of $L_x$
for $3 \le h \le 30$ if $L_x/d_s > 40$. Thus we have chosen $N$
for each simulation to guarantee $L_x/d_s \approx 50$.

In 3D, our system consists of a binary mixture of spheres
containing an equal number $N/2$ large spheres of diameter
$d_l$ and small spheres of diameter $d_s$ with a size ratio
$\sigma = d_l/d_s = 1.4$. Spheres are packed into a box of
dimensions $L_x$ by $L_y$ by $L_z$, with periodic boundaries
along the $x$- and $z$-directions and a fixed hard boundary
along the $y$-direction. Each configuration is generated
using the same particle expansion and contraction method
described above and the same initial values for $\delta{}\phi$
and the terminating conditions. For each configuration $L_x =
L_z$, $h = L_y/d_s$, and $N$ is chosen so that $L_x/d_s > 10$.
Our choice of $L_x/d_s > 10$ is not large enough to avoid finite
effects. However, in order to acquire the large amount of data
needed in a reasonable amount of time we intentionally choose
a value of $L_x/d_s$ below the finite size threshold. Trends
observed in the 2D analysis will be used to support that any
similar trends seen in 3D are real and not the result of the
finite periodic dimensions.  Note that in 3D we will show cases
where $h > L_x/d_s$ resulting in the confining direction being
larger than the periodic direction, and this may affect the
structure of final configurations; however, we will not draw
significant conclusions from those data.

Overall, it is not known if this algorithm produces
mathematically rigorously defined random close packed
states \cite{TorquatoPRL2000,XuPRE2005, Donev2004_2, OHern2004}.  However,
the goal of this article is to determine empirically the properties
of close-packed states in confinement, and we are not attempting to
extract mathematically rigorous results.  For example, we are not as
interested in the specific numerical values of $\phi_{rcp}$ that we
obtain, but rather the qualitative dependence on $h$.  As noted in
the introduction, different computational and experimental methods
for creating $rcp$ systems have different outcomes, and so it is
our qualitative results we expect will have the most relevance.

Note that for the remainder of this paper, we will drop the
subscript $rcp$, and it should be understood that discussions
of $\phi$ refer to the final state found in each simulation run,
$\phi_{rcp}(h)$.

\section{Results \label{section:Results}}

\begin{figure}
\includegraphics[width=3.4in]{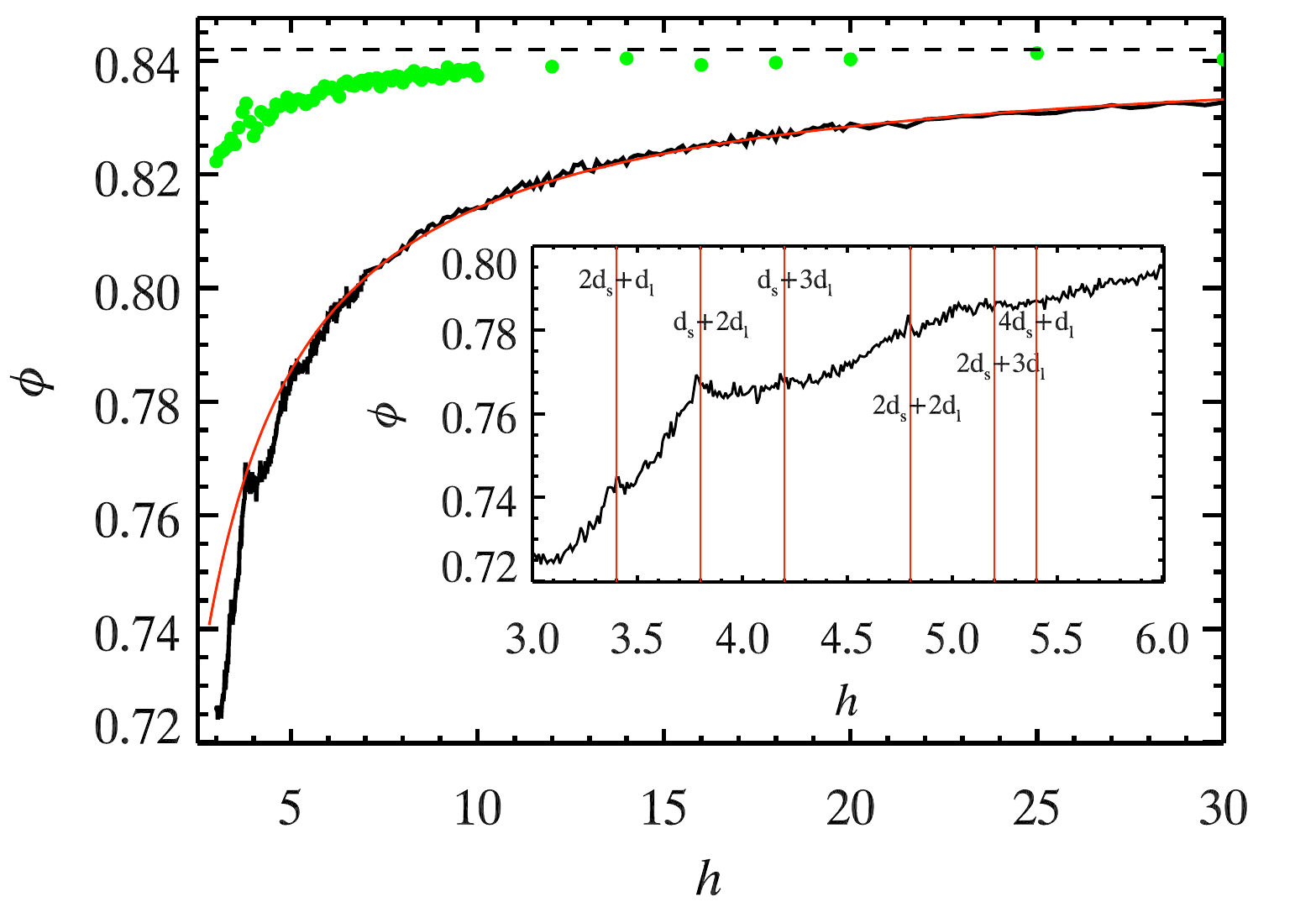}
\caption{(Color online). The black curve is the average packing
fraction $\phi$ found by averaging at least 10 2D configurations
together for various confining widths $h$; recall that $h$ has
been nondimensionalized by $d_s$, the small particle diameter. The
red curve (dark gray) is a fit using Eqn.~\ref{eq:model} which
finds $\phi_{rcp} = 0.842$ in the limit $h \rightarrow \infty$;
the value for $\phi_{rcp}$ is indicated by the black dashed line.
The green (light gray) data points are $\phi(h)$ computed for many
configurations with the confining wall replaced by a periodic
boundary.  The inset is a magnified view of the region for $h
\le 6$ to better show the large variations within this range. The
vertical lines in the inset are located at ``special" $h$ values
where peaks and plateaus appear.
}
\label{fig:2Dphivsh}
\end{figure}

\subsection{2D Systems}

We begin by generating many 2D configurations with $h$ between
3 - 30 and computing the packing fraction for each, as shown by
the black curve in Fig.~\ref{fig:2Dphivsh}.  This plot shows that
confinement lowers $\phi$, with the influence of the walls being
increasingly important at lower $h$.  The lowering of $\phi$ with
confinement is most likely due structural changes in the packing
near the confining boundary. We know that any alteration in particle
structure from a $rcp$ state must be ``near" the wall because as
$h \rightarrow \infty$ we expect to recover a packing fraction of
$\phi_{rcp}$, implying that in the infinite system the ``middle"
of the sample is composed of an $rcp$ region. Extrapolating the
data in Fig.~\ref{fig:2Dphivsh} to $h \rightarrow \infty$ we
find $\phi_{h \rightarrow \infty} = \phi_{rcp} = 0.842$ which is
essentially a test of our method. The extrapolation
(red curve in Fig.~\ref{fig:2Dphivsh}) was carried out by assuming
that to first order $\phi \sim \phi_{h \rightarrow \infty} - C/h$
for large $h$, where $\phi_{h \rightarrow \infty} = \phi_{rcp}$
(the bulk value for the $rcp$ packing)
and $C$ is a fitting parameter.

The data in Fig.~\ref{fig:2Dphivsh} begin to deviate from the fit
for $h \alt 6$, and furthermore $\phi(h)$ is not monotonic.  While some of
the variability is simply noise due to the finite number of disks
$N$ used in each simulation, some of the variability is real.
The inset in Fig.~\ref{fig:2Dphivsh} shows a magnified view of
the region $3 \le h \le 6$. The vertical lines in this inset are
located at specific values of $h$ that can be expressed as the
integer sums of the two particle diameters. For instance, the
first vertical line near the $y$-axis is located at $h = 2d_s +
d_l$. These lines are placed at some $h$ values where $\phi(h)$ has
notable spikes or plateaus.  These lines suggest that there exist
special values of $h$ where the confining thickness is the right
width so that particles can pack either much more efficiently or
much less efficiently than nearby values of $h$.  Intriguingly,
these special $h$ values are not the set of all possible integer
sums, but instead only the selected few drawn in the figure. For
example, there is no apparent feature near $h = 4$ (recall that $h$
is nondimensionalized by $d_s$).  Somewhat surprisingly, the peaks
correspond to integer combinations of $d_s$ and $d_l$, rather than
combinations such as $\sqrt{3}/2 d_s$.  The latter would suggest
hexagonal packing, the easiest packing of monodisperse disks in 2D;
whereas the observed peaks of $\phi(h)$ suggest square-like packing.

\begin{figure}
\includegraphics[width=3.4in]{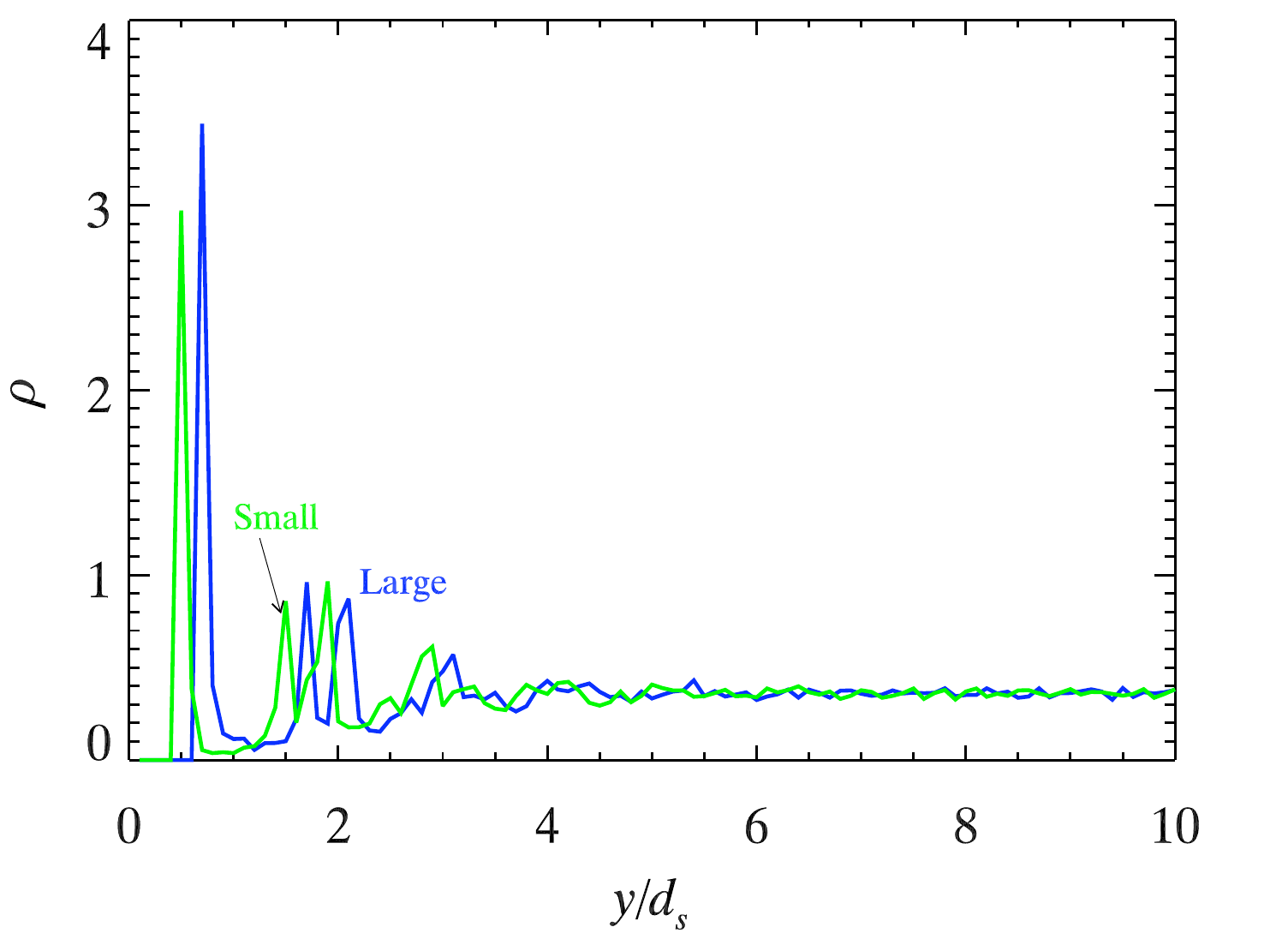}
\caption{A plot of the number density $\rho{}(y)$ for 100
2D configurations at $h = 30$ averaged together. The plot is
constructed by treating the small and big particles separately
and using bins along the confining direction of width $\delta{}y =
0.1d_s$.}
\label{fig:2Dsinglehistplot}
\end{figure}

To measure structural changes in particle packing as a result
of confinement we start by examining the variations in the
local number density $\rho$ with distance $y$ from the confining
wall. Figure \ref{fig:2Dsinglehistplot} is a plot of $\rho{}(y)$ for
100 configurations averaged together at $h=30$. This plot shows
oscillations in particle density which decay to a plateau. The
oscillations near the wall are indicative of particles layering
in bands. Above $y \gtrsim 6d_s$, noise masks these oscillations.
This supports our interpretation, that confinement
modifies the structure near the walls but not in the interior.
Furthermore, the rapidity of the decay to the plateau seen in
Fig.~\ref{fig:2Dsinglehistplot} suggests that confinement is only a
slight perturbation to systems with overall size $h \agt 6$.

The details of the density profiles in
Fig.~\ref{fig:2Dsinglehistplot} also suggest how particles pack
near the wall.  The small particle density (solid line) has an
initial peak at $y = 0.5 d_s$, indicating a large amount of small
particles in contact with the wall, as their centers are one radius
away from $y=0$.  Likewise, the large particle density (dashed
line) has its initial peak at $y = 0.7 d_s = 0.5 d_l$, indicating
that those particles are also in contact with the wall.  This is
consistent with the pictures shown in Fig.~\ref{fig:Configs}(a,b),
where it is clear that particles pack closely against the
walls.  Examining again the small particle number density in
Fig.~\ref{fig:2Dsinglehistplot} (solid line), the secondary peaks
occur at $y = 1.5 d_s$ and $y = 1.9 d_s = 0.5 d_s + 1.0 d_l$,
which is to say either one small particle diameter or one large
particle diameter further away from the first density peak at
$y=0.5 d_s$.  This again is consistent with particles packing
diameter-to-diameter, rather than ``nesting'' into hexagonally
packed regions.  Similar results are seen for the large particles
(dashed line) which have secondary peaks at $y = 1.0 d_s + 0.5 d_l$
and $y = 2.1 d_s = 1.5 d_l$.

\begin{figure}
\includegraphics[width=3.4in]{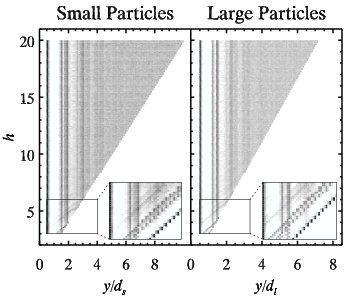}
\caption{An image representation constructed for the purpose of
comparing 2D $\rho(y)$ distributions at many different $h$. The
intensities have been logarithmically scaled. The vertical pixel
width is 0.1 and for the left plot the horizontal pixel width is
$0.2$ and for the right plot the horizontal pixel width is $0.14$.}
\label{fig:2Dhistpic}
\end{figure}

To confirm that these density profile results apply for a variety
of thicknesses $h$, and more importantly to see how these results
are modified for very small $h$, we use an image representation
shown in Fig.~\ref{fig:2Dhistpic}. To create this image, density
distributions of different $h$ are each separately rescaled to a
maximum value of 1. Every data point within each distribution is
then made into a gray scale pixel indicating its relative value;
black is a relative value of 1, and white is a relative value of
0. The vertical axis is the confining width and the horizontal axis
is the distance $y$ from the bottom wall. Each horizontal slice
(constant $h$) is essentially the same sort of distribution shown
in Fig.~\ref{fig:2Dsinglehistplot}. The white space on the right
side of the figure arises because the distribution is only plotted
for the range $0 \le y \le h/2$. The distributions are symmetric
about $y = h/2$, and by averaging the distribution found for the
range $0 \le y \le h/2$ with the distribution found for the range
$h/2 \le y \le h$, the statistics are doubled. The area shown in
the box is a magnified view of that region where the full range
$0 \le y \le h$ is being shown.

In Fig.~\ref{fig:2Dhistpic} there are vertical strips of dark
areas, once again indicating that particles are forming layers.
The width of these strips widens and the intensity lessens
farther from the wall.  In each plot, the first vertical black
strip is sharply defined and located at one particle radius,
illustrating that small and big particles are in contact with
the wall. Finally, the location and width of each layer remains
essentially the same for different $h$, suggesting that layering
arises from a constraint imposed by the closest boundary.  Given
that the first layer of particles always packs against the wall,
this imposes a further constraint on how particles pack in the
nearby vicinity. The consistency in the location and width of the
second layer for all $h$ demonstrates that the constraint of the
first layer always produces a similar packing in the second layer,
essentially independent of $h$. Continuing this argument, each layer
imposes a weaker constraint on the formation of a successive layer,
allowing for the local packing to approach $rcp$ far from the wall.

In the magnified views of Fig.~\ref{fig:2Dhistpic}, the 
vertical dark lines show the layering of particles induced by the
left boundary and the angled dark lines show the layering of
particles induced by the right boundary. We see that for small $h$
these sets of lines overlap and intersect, meaning that there is a
strong influence from one boundary on the packing within the
layers produced by the other boundary.  This may
explain the variations seen in $\phi(h)$ for small $h$ in
Fig.~\ref{fig:2Dphivsh}.  In particular, it is clear that at certain
values of $h$, the layers due to one wall are coincident with
the layers due to the other wall, and this suggests why $\phi(h)$
has a higher value for that particular $h$.  Given that the layer
spacings correspond to integer combinations of $d_s$ and $d_l$, the
coincidence of layers from both walls will correspond to integer
combinations of $d_s$ and $d_l$, and this thus gives insight into
the peak positions shown in the inset of Fig.~\ref{fig:2Dphivsh}.

\begin{figure}
\includegraphics[width=3.4in]{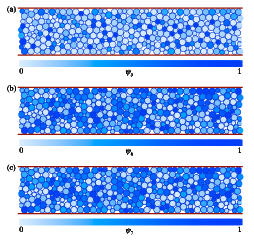}
\caption{(Color online). Drawings illustrating the conceptual
meaning of (a) $\psi_5$, (b) $\psi_6$, and (c) $\psi_7$.  Darker
colored particles have neighbors that are packed more like an ideal
regular n-sided polygon as compared to lighter drawn particles.
The configuration of particles is the same for all panels, and
are drawn from a simulation with $h=10$. Note in (b) that there
are no large patches of high $\psi_6$, demonstrating that there
are no large crystalline domains.
}
\label{fig:2Dpsiimages}
\end{figure}

\begin{figure}
\includegraphics[width=3.4in]{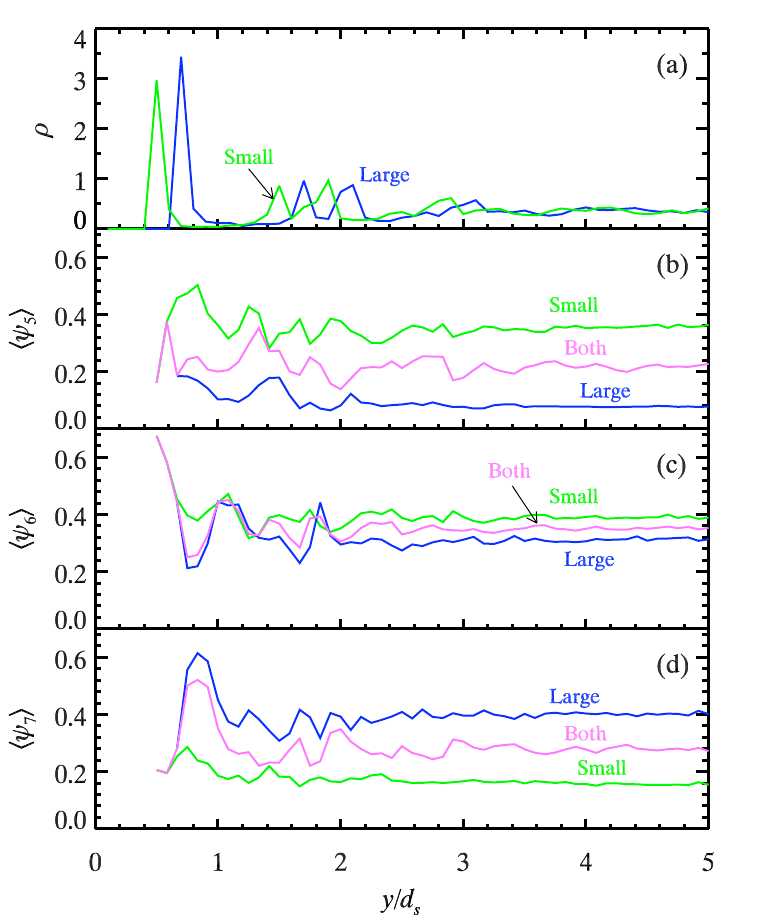}
\caption{ (Color online). (a) is a plot of the local number density
$\rho{}(y)$ for 2D configurations of big and small particles
separately.  (b) - (d) are plots of $\langle \psi_n \rangle(y)$
for  small (green/light gray) and big particles (blue/dark gray)
separately, and both sizes together (light purple/medium gray) where
(b) is $\langle \psi_5 \rangle$, (c) is $\langle \psi_6 \rangle$,
and (d) is $\langle \psi_7 \rangle$.  The length scales determined
from these curves for small, large, and both species are 
$\lambda_{5,s} = 1.2, \lambda_{5,l} = 1.1, \lambda_{5,b} = 1.4,
 \lambda_{6,s} = 0.8, \lambda_{6,l} = 0.9, \lambda_{6,b} = 0.8,
 \lambda_{7,s} = 1.1, \lambda_{7,l} = 0.7,$ and $\lambda_{7,b} = 1.0$
(all in terms of $d_s$).
}
\label{fig:2Dlengthscale}
\end{figure}


As described above, the influence of the walls diminishes rapidly
with distance $y$ away from the wall.  In particular, for the
local number density $\rho(y)$, we observe that the asymptotic
limit $\rho(y \rightarrow \infty) = 0.362$ for the curves shown
in Fig.~\ref{fig:2Dsinglehistplot} is in agreement with the
theoretical number density of an $rcp$ configuration $\rho_{rcp}
=  4\phi_{rcp}/\pi(1 + \sigma)$.  To quantify the approach to
the asymptotic limit, we define a length scale from a spatially
varying function $f(y)$ using:
\begin{equation}
\lambda = \frac{\int y\left[f(y) - f(y \rightarrow \infty)\right]^2
dy}{\int \left[f(y) - f(y \rightarrow \infty)\right]^2 dy}.
\label{eq:problengthscale}
\end{equation}
In this equation $f(y)$ is an arbitrary function where
the value of $\lambda$ quantifies the weighting of $f(y)$.
For simple exponential decay $f(y) = Ae^{(-y/\lambda')}$,
Eqn.~\ref{eq:problengthscale} gives $\lambda = \lambda'/2$.
Using $f(y) = \rho(y)$ we find $\lambda = 0.85 d_s$ and $\lambda =
0.72 d_s$ for the small particle curve and big particle curve in
Fig.~\ref{fig:2Dsinglehistplot} respectively, suggesting that the
transition from wall-influenced behavior to bulk $rcp$ packing
happens extremely rapidly.

To further investigate the convergence of the local packing to $rcp$
more closely we analyze the local bond order parameters $\psi_n$,
which for a disk with center of mass $r_i$ are defined as
\begin{equation}
\psi_n(r_i) = \frac{1}{n_b} \displaystyle\sum_j e^{ni\theta(r_{ij})}.
\end{equation}
The sum is taken over all $j$ particles that are neighbors of
the $i$th particle, $\theta(r_{ij})$ is the angle between the
bond connecting particles $i$ and $j$ and an arbitrary fixed
reference axis, and $n_b$ is the total number of $i$ - $j$
bonds~\cite{MarcusPRL1996}. The magnitude of $\psi_n$ is bounded
between zero and one; the closer the magnitude of $\psi_n$ is to 1,
the closer the local arrangement of neighboring particles are to
an ideal $n$-sided polygon.  Figures~\ref{fig:2Dpsiimages}(a-c)
are drawings illustrating the concept of $\psi_n$ using a 2D
configuration with $h = 10$. Particles with larger $\psi_n$ are
drawn darker.  These figures have no large clusters of dark
colored particles, demonstrating that there are no large crystalline
domains (i.e. particles are randomly packed).

For a highly ordered monodisperse packing $\langle \psi_6 \rangle$
would be the most appropriate choice for measuring order because
of the ability for monodisperse packings to form hexagonal
packing. However for a bidisperse packing with size ratio $\sigma =
1.4$, the average number of neighbors a small particle will have
is 5.5 and the average number of neighbors big particles will have
is 6.5. Therefore, a bidisperse packing of this kind will have
a propensity to form local pentagonal, hexagonal, and heptagonal
packing, and to properly investigate how the local packing varies
we examine $\langle \psi_5 \rangle$, $\langle \psi_6
\rangle$, and $\langle \psi_7 \rangle$.
We compute the average values $\langle \psi_5 \rangle$,
$\langle \psi_6 \rangle$, and $\langle \psi_7 \rangle$ for all
configurations as a function of $y$, and averaging together
all $\langle \psi_n \rangle$ distributions for configurations
with $h \ge 16$ to improve statistics. This averaging can
be justified by considering that oscillations in $\rho(y)$ in
Fig.~\ref{fig:2Dsinglehistplot} for $y/d_s > 10$ are quite small.
Thus this averaging improves our statistics for the range $0 <
y/d_s < 5$ where the largest oscillations occur, without skewing the
data. In the end nearly 10,000 configurations are averaged
together, producing the curves shown in
Fig.~\ref{fig:2Dlengthscale}(b-d).  This figure
shows the spatial variations of $\langle
\psi_5 \rangle$, $\langle \psi_6 \rangle$, and $\langle \psi_7
\rangle$ for small and big particles separately and both
particles combined. All curves show fluctuations that decay
with distance from the wall, and show local order within and
between layers. Figure \ref{fig:2Dlengthscale}(a) has been added so
comparisons between the location of the oscillations in $\rho(y)$
and $\langle \psi_n\rangle(y)$ can be made.

For the most part, each successive layer has less orientational
order than the previous layer with $\langle \psi_n\rangle$
eventually decaying to an asymptotic limit. To characterize
a length scale for these curves we compute $\lambda$
using Eqn.~\ref{eq:problengthscale} for each curve shown in
Fig~\ref{fig:2Dlengthscale}(b-d).  From the nine curves, we find
that the mean value of $\lambda = (1.00 \pm 0.24) d_s$. The length scales
found for these curves are once again less than the largest particle
diameter.  No striking difference is found between the different
order parameters or between the different particle sizes; specific
values of $\lambda$ are given in the figure caption.  (Note that
the asymptotic limits of all $\langle \psi_n \rangle$ plots are in
agreement with the average values found for 40 unconfined 10,000
particle simulations averaged together, confirming that the local
packing converges to an $rcp$ arrangement far from the walls.)

\begin{figure}
\includegraphics[width=3.4in]{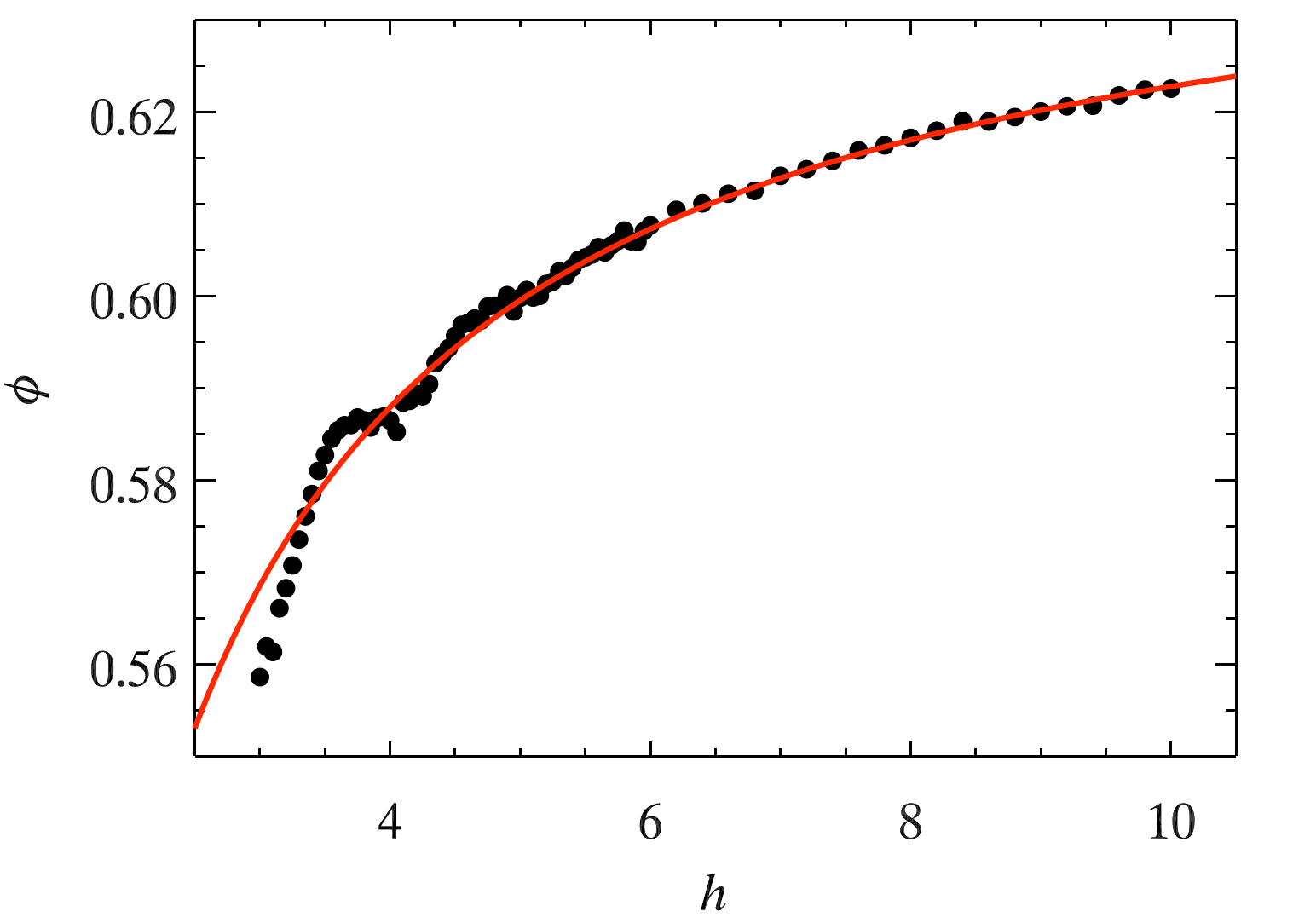}
\caption{The black data points are the average packing fractions of
3D configurations at various $h$. The red (dark gray) curve is a
fit from our model Eqn.~\ref{eq:model}. For each $h$ at least 10 configurations were averaged together.}
\label{fig:3Dphivsh}
\end{figure}

Next, we wish to distinguish the structural influence of the flat
wall from the finite size effects.  We perform simulations
where the confining wall is replaced by a periodic boundary with
periodicity $h$; thus particles cannot form layers.  In this case,
the packing fraction still decreases as $h$ is decreased, as
shown by the green curve (light gray) in Fig.~\ref{fig:2Dphivsh},
although the effect is less striking than for the case with walls
(black curve).  A likely explanation for the decrease in $\phi$
with confinement is the long range structural correlations imposed
along the constricted direction; in other words, if there is a
particle located at $(x,y)$ that particle is mirrored at $(x,y-h)$
and $(x,y+h)$ by the periodic boundary.  We know from the pair
correlation function~\cite{Lochmann2006,KansalChemPhys2002} of $rcp$
configurations that structural correlations exist over distances
of many particle diameters, although of course these are weak
at larger distances.  Thus the periodicity forces a deviation
from the ideal $rcp$ packing, that becomes more significant as
$h$ decreases.  By definition $rcp$ is the most random densely
packed state, and thus any perturbations away from this state
must have a lower packing fractions.  However, this is not nearly
as significant as the constraint imposed by the flat wall, as is
clear comparing the green (light gray) data and the black data
in Fig.~\ref{fig:2Dphivsh}.

\subsection{3D Systems}

We next consider 3D confined systems.  We start by investigating
$\phi(h)$, shown as the black points in Fig.~\ref{fig:3Dphivsh}. As
observed in the 2D case, $\phi$ is reduced as a result of
confinement. However, unlike the 2D system, there does not appear
to be a series of ``special values" of $h$ that give rise to peaks
and plateaus, other than a hump near $h = 3.75$.  The lack of
substructure may be due to the smaller size in $x$ and $z$, in
contrast with the 2D simulations which had large sizes in the
unconfined direction.

\begin{figure}
\includegraphics[width=3.4in]{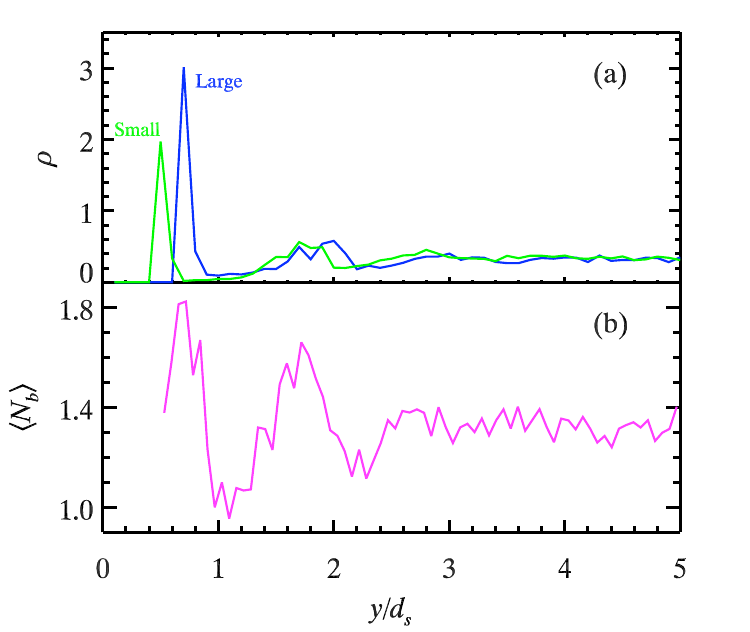}
\caption{(a) is plot of $\rho(y)$ for 3D
configurations for small and big particles separately. The plot
is constructed using bins of width $\delta{}y = 0.1d_s$ along the
confining directions. 
(b) is a plot of the average number of
ordered bonds $\langle N_b \rangle(y)$.}
\label{fig:3Dlengthscales}
\end{figure}

Next we investigate the local number density $\rho(y)$ for
$h = 25$, shown in Fig.~\ref{fig:3Dlengthscales}(a). The data
are constructed by averaging together 100 configurations. The
curve shows fluctuations that decay with distance from the wall,
eventually reaching a plateau.  Using Eqn.~\ref{eq:problengthscale},
we obtain decay lengths $\lambda_{3D} = 0.77d_s$ and $0.73d_s$
for the small and large particle curves respectively. These length
scales are similar to the length scales obtained in the 2D case
($\lambda_{2D} = 0.85d_s$ and $0.72d_s$ for small and large
particles). 

\begin{figure}
\includegraphics[width=3.4in]{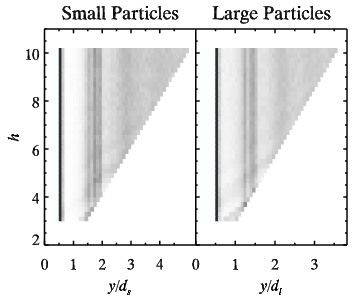}
\caption{An image representation comparing the number density distributions of 3D configurations for many different $h$. Black pixels represents a relative value of 1 and white represent a relative value of 0. A gray scale is used to represent relative values between 0 and 1. The pixel widths are 0.1$d_s$ horizontally and 0.2 vertically.}
\label{fig:3Dhistpic}
\end{figure}

To compare all 3D $\rho(y)$ distributions for different $h$
we construct the image representation used to compare 2D
configurations in Fig.~\ref{fig:2Dhistpic}.  The data for the 3D
configurations are shown in Fig.~\ref{fig:3Dhistpic}.  Again there
are dark vertical strips arising from particles forming layers
near the wall.  Like in 2D, the density approaches the ``bulk''
$rcp$ value far from the wall.

In 2D, we also noted that the structure is modified near the wall,
as measured by the $\psi_n$ order parameters.  
To investigate structural ordering in 3D, we use a local structural
parameter sensitive to ordering 
\cite{Steinhardt1983,Gasser2000}.  We start by defining
\begin{equation}
\hat{q}_{i,6}  = \frac{1}{n_j K} \displaystyle\sum_{j}
Y_{6m}(\theta_{ij}, \phi_{ij}).
\label{eq:q6}
\end{equation}
In the above equation $m = \{-6, ..., 0, ..., 6\}$, and
thus $\hat{q}_{i,6}$ is a 13 element complex vector which
is assigned to every particle $i$ in the system. The sum in
Eqn.~\ref{eq:q6} is taken over the $j$ nearest neighbors of the
$i$th particle, $n_j$ is the total number of neighbors, and $K$ is
a normalization constant so that $\hat{q}_{i,6} \cdot \hat{q}_{i,6}
= 1$. For two particles $i$ and $j$ that are nearest neighbors,
$Y_{6m}(\theta_{ij},\phi_{ij})$ is the spherical harmonic associated
with the vector pointing from particle $i$ to particle $j$, using
the angles $\theta_{ij}$ and $\phi_{ij}$ of this vector relative to
a fixed axis.  Next, any two particles $m$ and $n$ are considered
``ordered neighbors'' if $\hat{q}_{m,6} \cdot \hat{q}_{n,6} > 0.5$
\cite{Steinhardt1983,Gasser2000}.  Finally, we quantify the local
order within the system by the number of ordered neighbors $N_b$ a
particle has.

Figure~\ref{fig:3Dlengthscales}(b) is a plot of the average
number of ordered neighbors particles have $\langle N_b \rangle$
as a function of distance $y$ from the wall.  In comparison with
Fig.~\ref{fig:3Dlengthscales}(a), this plot shows that local order
is mostly seen within layers, not between layers. Also we see
that $\langle N_b \rangle$ converges to an asymptotic value of
$\approx 1.3$, confirming that the system is disordered (values
of $N_b > 8$ are considered crystalline \cite{Gasser2000}).
We use Eqn.~\ref{eq:problengthscale} to characterize a length
scale for the decay in $\langle N_b \rangle$, giving $\lambda
= 1.3d_s$.  The asymptotic limit of $\langle N_b \rangle(y)$ in
Fig.~\ref{fig:3Dlengthscales}(b) agrees with the average value
of $N_b$ found for 15 large simulations with 2,500 particles and
periodic boundary conditions, confirming that the local structure in
the confined case converges to the bulk $rcp$ state far from
the walls.

Our results show that in both 2D and 3D, confinement induces
changes in structural quantities near the walls, with a decay
towards the ``bulk'' values characterized by length scales no
larger than $d_l$.  The only prior work we are aware of with
related results are a computational study~\cite{LaundryPRE2003}
and an experimental study~\cite{SeidlerPRE2000} of collections of
monodisperse particles confined in a large silo.  The simulation
by Landry et al.~primarily focused on the force network within
the silo.  They show one plot of the local packing fraction
as a function of distance from the silo wall.  Similar to our
results, this local packing fraction showed fluctuations that
decayed monotonically. In their paper they state a decay length
of $\approx 4d_l$; however, it appears that they drew this
conclusion by estimating the value by eye.  Applying Eqn.~\ref{eq:problengthscale} to their data we find $\lambda$ on the order of $d_l$, close to the value found in our simulations. The experimental study
by Seidler et al. reported on the local bond orientational order
parameter which showed oscillation that decayed with distance from
the wall. They reported a decay length of $\lambda \approx d_l$
using an exponential fit. The length scales from these two studies
are slightly larger than those found in our work.

\section{Model \label{section:Discussion}}

\begin{figure}
\includegraphics[width=3.4in]{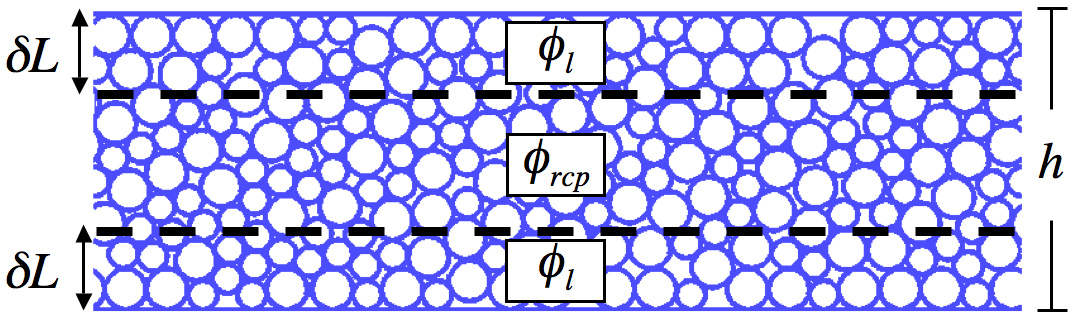}
\caption{(Color online). Illustration of our model for $\phi(h)$.
Our model breaks a configuration with confining width $h$ into
three regions.  The boundary layers are approximated to have a
packing fraction $\phi_l$ and persist a distance $\delta{}L$ into
the sample, and the middle ``bulk" region is approximated to have
a packing fraction $\phi_{rcp}$.}
\label{fig:model}
\end{figure}

Our results for $\phi(h)$ can be understood with a
simple model incorporating an effective boundary
layer and a bulk like region.
Figure~\ref{fig:model} shows a configuration of particles
confined between two plates and divided into three regions. Near
each confining wall particles show strong layering and will assume
a configuration much different than $rcp$. For this model, these
particles will be replaced by an effective ``boundary layer" with
packing fraction $\phi_l$. Layering will persist far into the bulk,
but after some distance, indicated in the figure by $\delta L$, the
local packing of particles will be near that of $rcp$. Particles in
the bulk region will be approximated to be in a $rcp$ configuration
with packing fraction $\phi_{rcp}$. Using this simple model,
$\phi$ can be approximated by the weighted average $\phi = \frac{h
- 2\delta{}L}{h}\phi_{rcp} + \frac{2\delta{}L}{h}\phi_{l}$ (in
either 2D or 3D, with of course different values of the parameters
depending on the dimension.)
Reducing this equation further we obtain
\begin{equation} \phi =
\phi_{rcp} - \frac{C}{h}, \label{eq:model} 
\end{equation}
where $C = 2\delta{}L(\phi_{rcp} - \phi_l)$.  Note that this is
the same form for $\phi(h)$ obtained from considering a 1st order
correction in terms of $1/h$.

\begin{figure}
\includegraphics[width=3.4in]{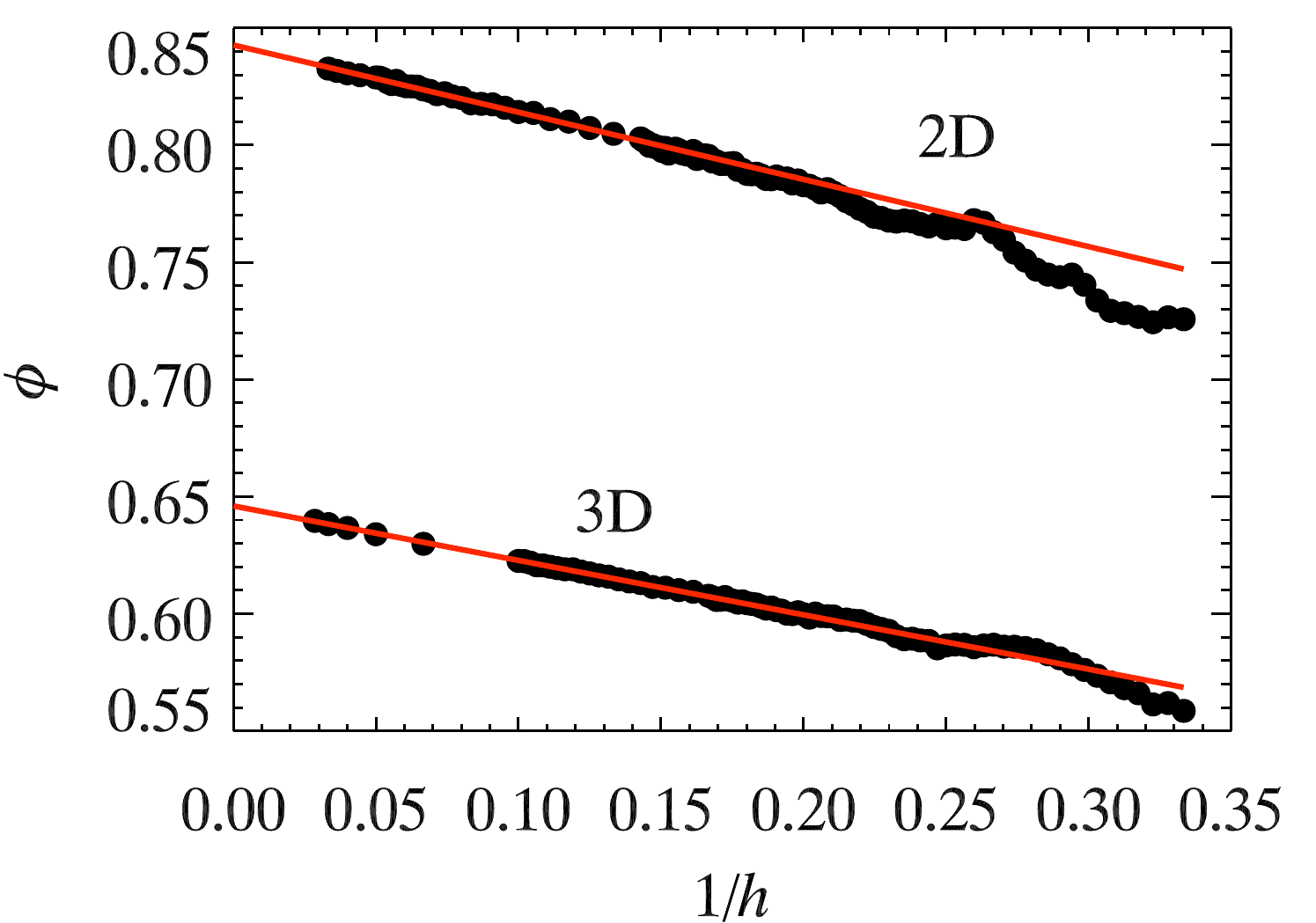} 
\caption{(Color
online). The upper black curve is a plot of $\phi(1/h)$ for 2D
configurations, and the red (dark gray) line going through the
curve, is a fit from our model. Likewise, the lower black curve is
a plot of $\phi(1/h)$ for 3D configurations with the red (dark gray)
line going through the curve being another fit from our model.}
\label{fig:modelfit}
\end{figure}

Equation~\ref{eq:model} only contains two fitting parameters,
$\phi_{rcp}$ and $C$. $\phi_{rcp}$ is the packing fraction of
an infinite unconfined configuration, and $C$ approximates
the difference in $\phi$ from $\phi_{rcp}$ for the first
layers contained within a distance $\delta{}L$ from the
confining wall. The data in both Fig.~\ref{fig:2Dphivsh} and
Fig.~\ref{fig:3Dphivsh} are fitted to Eqn.~\ref{eq:model}. The fits
are shown as the red lines (dark gray) in the earlier figures,
and also in Fig.~\ref{fig:modelfit}, where the data are plotted as
functions of $1/h$ to better illustrate the success of this model.
The fits give for 2D $\phi_{rcp} = 0.844$ and $C = 0.317$ and for
3D $\phi_{rcp} = 0.646$ and $C = 0.233$. Both fits give values
for $\phi_{rcp}$ that are slightly larger, but not by much, than
$\phi_{rcp}$ reported earlier in the paper.  In Fig.~\ref{fig:modelfit} is can be seen
that the packing fraction for large $1/h$ dip significantly below
the fitting line, due to the fluctuations in $\phi(h)$ for small
$h$; this is responsible for the over estimate in $\phi_{rcp}$.
When the data for both curves are fitted for $h \ge 8$ ($1/h <
0.125$) the actual values for $\phi_{rcp}$ are obtained.

\begin{figure}
\includegraphics[width=3.4in]{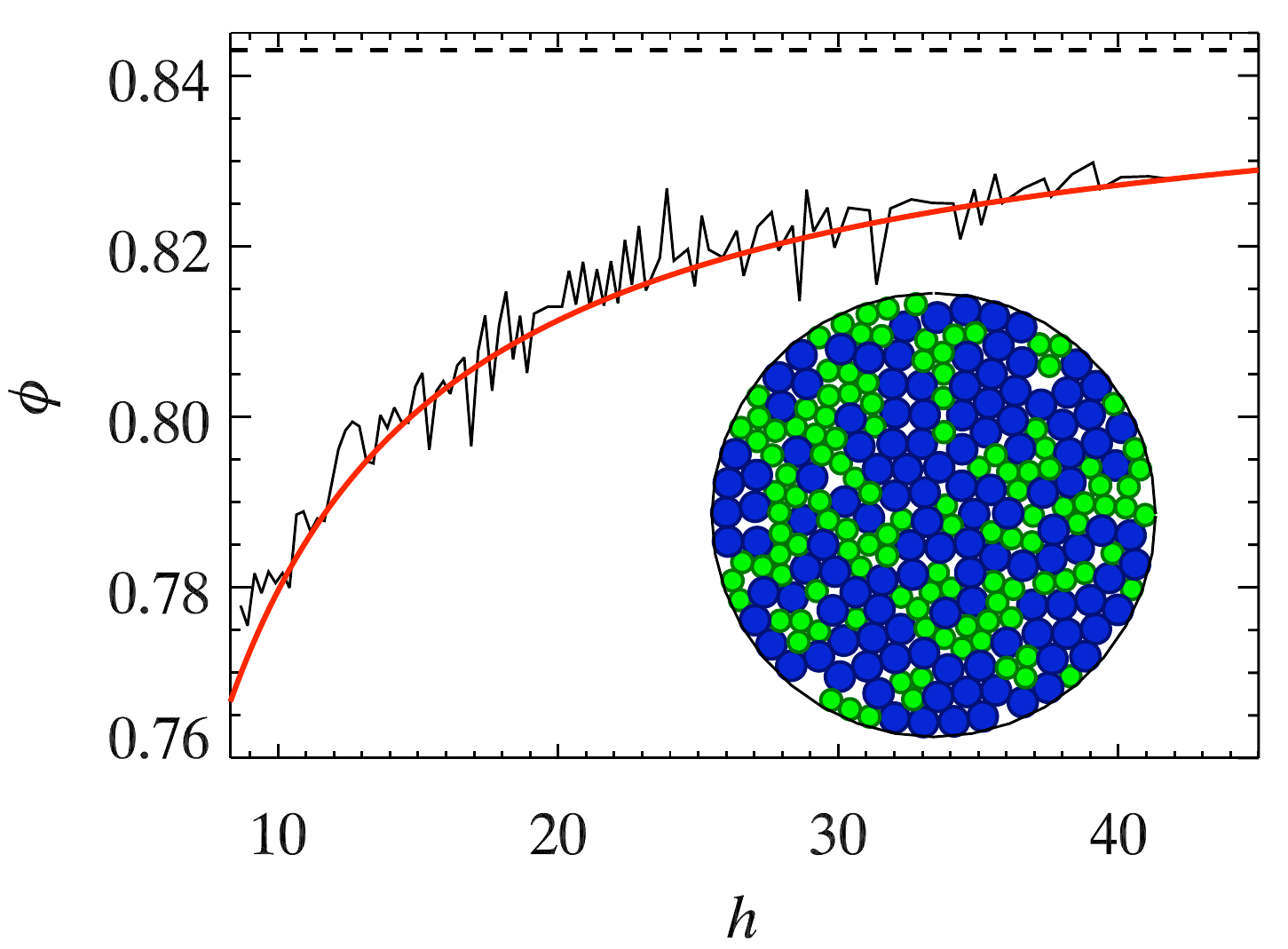} 
\caption{(Color online). The black curve is the packing fraction
dependence of random close packed 2D disks enclosed within
a circular boundary. The red (light gray) curve is from our
model using the parameters appropriate to Fig. 12 (2D case).
The image at the lower right is an rcp configuration confined
in a circular boundary with $h = 21$. Small particles are rendered
as green (medium gray) and large particles are rendered as blue
(dark gray).}
\label{fig:circularboundary}
\end{figure}

To provide further credence to our model we also perform 2D $rcp$
simulations with a fixed circular boundary.  Our simulations
were carried out with different confining widths ranging from
$h \approx 10 - 40$, where $h$ is the diameter of the circular
boundary normalized by $d_s$. Figure~\ref{fig:circularboundary}
shows a plot of $\phi(h)$ for a circular fixed boundary. As before
with a flat boundary condition, we see that $\phi$ increases to
an asymptotic limit.  Adapting our model to a curved boundary with
radius $R=h/2$, the weighted average becomes $\phi(R) = \frac{\pi
[R^2 - (R-\delta L)^2]}{\pi R^2}\phi_{rcp} + \frac{\pi (R-\delta
L)^2}{\pi R^2}\phi_{l}$.  This expression can be simplified as
$\phi = \phi_{rcp} - 2C/h$, using $C=2 \delta L (\phi_rcp-\phi_l)$
as before, and dropping a term that is second order in $\delta L
/ h$.  We show this curve as the red line (light gray) curve in
Fig.~\ref{fig:circularboundary}, using the same values from our
prior 2D fit with non-curved walls ($\phi_{rcp} = 0.844$, $C =
0.317$), and find good agreement with the data.  A direct fit to
the circular data gives $\phi_{rcp} = 0.843$ and $C = 0.301$; the
slightly smaller value of $C$ suggests that particles pack more
efficiently near a curved boundary than a flat boundary.  However,
within the uncertainty of our limited data in the curved geometry,
it is not clear if the difference in $C$ is significant.

\section{Conclusion \label{section:Conclusion}}

In this paper, we have shown how a confining boundary alters
the structure of random close packing by investigating simulated
$rcp$ configurations confined between two walls in 2D and 3D. We
find that confinement lowers the packing fraction, and induces
heterogeneity in particle density where particles layer in bands
near the wall. The structure of the local packing decays from a
more ordered packing near the wall to a less ordered packing in
the bulk.  All measures of local order and local density decay
rapidly to their bulk values with characteristic length scales on the
order of particle diameters.  Thus, the influence of the walls is
rapidly forgotten in the interior of the sample, with
confinement having the most notable effects when the confining
dimension is quite small, perhaps less than 10 particle diameters
across.

These findings have implications for experiments investigating
the dynamics of densely packed confined systems (i.e. colloidal
suspensions or granular materials).  For example, our work
shows that for small $h$ the packing fraction has significant
variations at small $h$ (mostly clearly seen in 2D, for example
Fig.~\ref{fig:2Dphivsh}).  For dense particulate suspensions with
$\phi < \phi_{rcp}$, flow is already difficult.  By choosing a value
of $h$ with a local maximum in $\phi_{rcp}(h)$, a suspension may be
better able to flow, as there will be more free volume available.
Likewise, a poor choice of $h$ may lead to poor packing and
enhanced clogging.  A microfluidic system with a tunable size
$h$ may be able to vary the flow properties significantly with
small changes of $h$, but our work implies that control over $h$
needs to be fairly careful to observe these effects.  Of course,
these effects will be obscured by polydispersity in many systems of
practical interest; however, our work certainly has implications
for microfluidic flows of these sorts of materials, once the
minimum length scales approach the mean particle size.

Our work has additional implications for experiments on confined
glasses \cite{NugentPRL2007, Morineau2002, Schuller1994,
Jackson1991, Kim2000, Thompson1992, Nemeth1999, Scheidler2002,
Alcoutlabi2005}.  As mentioned in the introduction, confinement
changes the properties of glassy samples, but it is unclear if this
is due to finite size effects or due to interfacial influences
from the confining boundaries \cite{He2007}.  Our results show
that dense packings have significant structural changes near the
flat walls, suggesting that indeed interfacial influences on
materials can be quite strong at very short distances, assuming
that the structural changes couple with dynamical behavior.

\begin{acknowledgments}
This work was supported by the National Science Foundation under
Grant No.~DMR-0804174.
\end{acknowledgments}


\end{document}